# AI-Driven SERS for Non-invasive and Label-Free Extracellular Vesicle Detection Across Cellular Origins in Tears and Sweat

*Yang Li*[a,b,c,1,*], *Xiaoming Lyu*[b,1], Ling Xia[b], *Kuo Zhan*[a], *Haoyu Ji*[d], *Lei Qin*[b], *Seppo J. Vainio*[e], *Jian-An Huang*[a,e,*]

a. Research Unit of Health Sciences and Technology (HST), Faculty of Medicine, University of Oulu, Finland

b. Research Center for Innovative Technology of Pharmaceutical Analysis, College of Pharmacy, Harbin Medical University, Heilongjiang 150081, PR China

c. National Key Laboratory of Frigid Zone Cardiovascular Diseases (NKLFZCD), College of Pharmacy, Harbin Medical University, Heilongjiang 150081, PR China

d. Department of Pharmacy at The Second Affiliated Hospital, and Department of Pharmacology at College of Pharmacy (The Key Laboratory of Cardiovascular Medicine Research, Ministry of Education), Harbin Medical University, Harbin 150081, PR China

e. Research Unit of Disease Network, Faculty of Biochemistry and Molecular Medicine, University of Oulu, Finland

* Corresponding authors:

Yang Li, email: liy@hrbmu.edu.cn, Yang.Li@oulu.fi.

Phone number: +86 173 4567 8922

Jian-An Huang, email: Jianan.Huang@oulu.fi.

[1] These authors contributed equally to this work.

## Abstract

Wearable sensing technology capable of point-of-care, continuous and non-invasive analysis of exosomes in biofluid such as tears and sweat is an essential part for future personalized medicine. Major detection and identification methods of cell secreted Extracellular Vesicles (EVs) often require labeling and are time-consuming, resulting in low efficiency in EV mechanism research and disease diagnosis. While the label-free Surface-enhanced Raman spectroscopy (SERS) has been combined with deep learning model for EV identification in blood, their application to non-invasive detection of EVs in tears and sweat are missing. Here, we filled this gap by developing an artificial intelligence (AI)-assisted Surface-enhanced Raman spectroscopy (SERS) method based on salt-induced nanoparticle aggregation for fast EV identification in tears and sweat with high accuracy. Significantly, our label-free detection and AI differentiation of EVs from 6 cell lines (HepG2, Hela, 143B, LO-2, BMSC, H8) achieved the identification of EVs in tear fluids from 7 different disease sources with accuracies >92%. Our results showed that this platform can not only distinguish EVs from multiple cell sources but also generate highly reproducible and selective EV signals in tear fluids without a need for chemical labeling or separation steps. Molecular dynamics simulations revealed that silver atoms (Ag) form electrostatic interactions with oxygen atoms of multiple amino acid residues in proteins, suggesting a high affinity. This strategy realizes ultra-sensitive and anti-interference detection of EVs, providing a new idea for the rapid diagnosis of clinical diseases.

# 1. Introduction

Cell Secreted Extracelular nano- or microsize molecularly loaded vesicles (Exosomes when derived from the multivesical bodies) are lipid and corona encircled vesicles that are actively assembled and secreted by most of not all living cells, with a diameter that typically is around 30-150 nm. EVs transmit nucleic acids, proteins, and lipids, and they are considered to function as an important communication mechanism between cells and have during recent years attracted much attention[1-3]. Initially, EVs were thought to be primarily carriers of cellular waste via human fluids, such as saliva, sweat, blood, and urine[4]. However, subsequent studies have started to provide evidence that the content, namely the specific molecular signature of EVs can also reflect the pathophysiological status of parental cells, so that EV loaded proteins, nucleic acids, and lipids composition can become altered in reference to changes in the health status of cells, tissues and organs [5]. Thus, EVs are expected to play important roles in intercellular communication, and diseases such as tumourigenesis [6-9].

Liquid biopsy methods for EVs where these are extracted from serum provide a promising diagnostic layer but represent still an invasive medical diagnostic approach but provide benefits to the biopsies . EVs now prvide yet additional expected benefits to diagnostics since they cargo wealth of candidate disease biomarkers including cancer [10-13].

Well use methods for EVs detection include transmission electron microscopy (TEM), nanoparticle tracking analysis (NTA), protein immunoblotting (Western blot), and enzyme-linked immunosorbent assay (ELISA) [14, 15]. However, these analytic tools are time-consuming, laborious and need specific EV based molecular recognition and the routine ones are of low resolution. As new technological openings to measure accurately such nanoscale particles methods as fluorescence-integrated microfluidics electrochemical sensors, and CRISPR/CAS system-assisted detection have been adoptred [16-18], These apprisches have improved the sensitivity and EV detection limit, but require still complicated EV pretreatment processes and are technically yet limited in their application potential.

Surface-enhanced Raman spectroscopy (SERS) has been widely used in the field of analytical detection due to its simple and fast measure and analyte detection, and as sample handling is relatively easy, and especially SERS offer high sensitivity, and nondestructiveness [19-23]. Due to these reasons SERS has been applied as a qualitative and quantitative EV analytic approach.

Typically EV detection is divided into two over all methods, either labelled or label-free[24]. Zong et al. fabricated anti-CD 63 antibody-modified magnetic nanoparticles and anti-HER 2 antibody-modified Au@Ag nanorods for qualitative and quantitative detection of EVs [25]. Pang et al. used anti-PD-Ll antibody-modified Au@Ag@MBA as a SERS tag for EV labelling with PD-Ll for quantitative purposes [26]. Han et al. in turn used EV protein biomarkers for the diagnosis of osteosarcoma associated EVs [27]. However the limitation of these studies is spesific target molecular tags, labels or antibodies are needed. To overcome the needs for EVs labels, researchers have developed label-free EV assays. As an example Lee et al. used a silver-film-coated nanobowl platform to capture EVs secreted by the SKOV 3 cells [28]. Stremersch and colleagues prepared positively charged gold nanoparticles by coating them with a 4-dimethyl aminopyridine layer to depict EVs from various cellular sources [29]. Dong et al. developed a gold-coated $TiO_2$ macroporous inverse opal (MIO) structure to capture and analyze plasma EVs of patients with cancer [30].

Machine Learning methods that are developing in connection to the Artificial Intelligence (AI) has been combined with SERS to improve EV based analysis. Algorithms such as Partial Least Squares Regression (PLS), Linear Discriminant Analysis (LDA), Random Forests (RF), Support Vector Machines (SVM), and Convolutional Neural Networks (CNN), Machine Learning serve to enable analysis of large data sets, spotted patterns and trends in data. Moreover the developed algoritms provide means to make better predictions from the generated data serving also as the ground for more efficient data interpretation. Since the Raman spectra as multivariate data provides sample approximations, machine learning algoritms will provide in combination with SERS more robust analytic power for the raman spectral capacity as a diagnostic measure [31-34]. For example Zhe Zhang et al. combined LDA to achieve better classification and identification of adenoviruses [35], Janina Kneipp et al. lipid analytics [36] and Hyunku Shin et al. apploied the approach to analyze SERS features of blood associated EVs to address early stage cancer diseases [37].

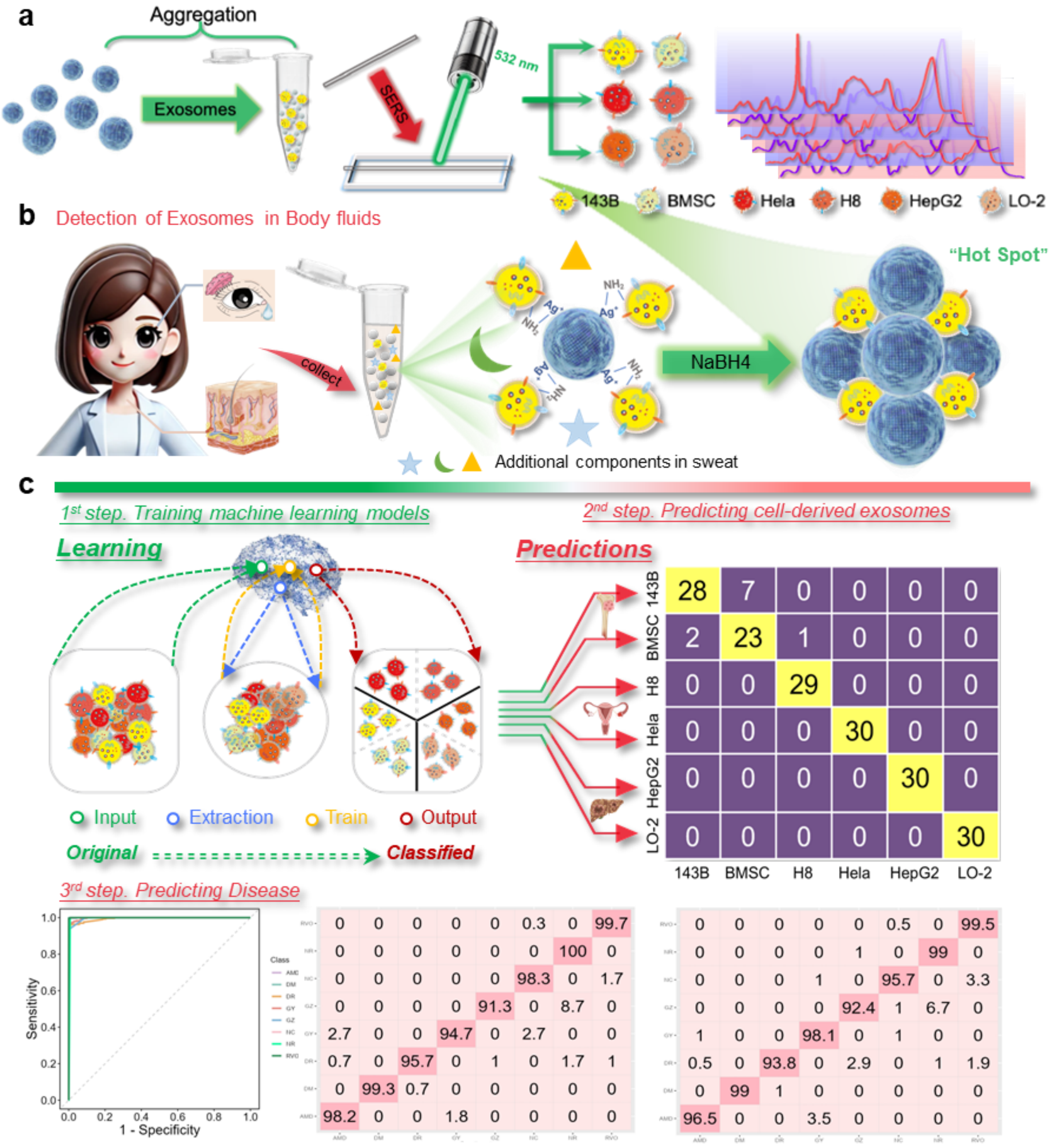

*Figure 1: Schematic illustration of SERS-based EV detection and machine learning-assisted differentiation of EVs from multiple sources. a. Detection process of exosomes derived from multiple cell types. b. Detection process of exosomes in body fluids, including tears and sweat. c. Machine learning-based differentiation of EVs from distinct cellular origins. In the first step, 70% of the SERS spectral data is used as the training set. Following feature extraction, a support vector machine (SVM) model is trained. In the second step, the remaining 30% of the data serves as the test set for prediction, and the classification performance is summarized in a confusion matrix. In the third step, the model further distinguishes among multiple ophthalmic diseases.*

In this work we targeted EVs that associate to sweat and tears adopting EV nanoparticle based clustering to make them visible for SERS based measure and AI machine learning applicaton We used label-free SERS to detect EVs from selected cellular sources and combined them with the PCA-SVM algorithm for differentiation and prediction **(Figure 1).** Compared to CNN, the SVM algorithm performs well with small sample datasets, decision boundaries are more precise, overfitting can be effectively avoided, and model results are easier to interpret, which has gained wider application[38-41].

We prepared silver nanoparticles by reducing silver nitrate with sodium borohydride as an enhance-substrate for SERS and used sodium borohydride solution as an aggregation agent for detecting EV signal. We obtained EV SERS spectra from three selected model cancer cells including normal cells that serves as controls. To further investigate the heterogeneity of EVs due to expected differences in their molecular composition depending of their origin, we applied PCA to downscale the SERS data. Here we combined the data with the SVM algorithm for analysis and prediction. Based on this method, EVs derived from tears from seven types of diseases were further detected. The convolutional neural network (CNN) algorithm was combined to realize the differentiation and prediction of diseases.

The results depicted that this method can label-free obtain fingerprint feature profiles of EVs from different sources conveniently and quickly, without the influence of EV concentration and identify them according to their nuances.

# 2. Materials and methods

## 2.1 Chemicals and reagents

DMEM high glucose medium was purchased from Gibco, F12 medium was purchased from Sigma-Aldrich, and all cell lines were purchased from American Type Culture Collection (ATCC, China). Silver nitrate ($AgNO_3$, ACS, 99.9%) was purchased from Alfa Aesar (China) Chemical Co. Ltd; sodium borohydride (99.99%) was purchased from Sigma-Aldrich Co. Ltd (Shanghai, China).

## 2.2 Cell culture

Hela, H8, HepG2, LO-2, and 143B cells were cultured in culture flasks using DMEM high-sugar medium supplemented with 10% fetal bovine serum (FBS) and 1 mmol penicillin/streptomycin. BMSC cells were cultured in culture flasks using F12 medium with 10% fetal bovine serum (FBS) and 1 mmol penicillin/streptomycin. All cells were grown at 37°C and 5% CO2, the medium was changed every 2-3 days, and the cells were passaged at 90% density. When EVs were collected, the passaged cells were cultured until confluent with the above-growth medium. The cells were then washed with PBS (3 times) and starved in medium prepared in the presence of serum-free.

## 2.3 EV (Exosomes) isolation and characterization

After 48 hours of starvation incubation, the cell supernatant was harvested for EV isolation, EVs were purified by routine ultracentrifugation, so that the supernatant was centrifuged at 300 × g for 10 min to remove cells, 3,000 × g for 20 min to remove cell dead cells, 10,000 × g for 30 min to remove cellular debris, followed by a centrifugation at 120,000 × g for 70 min to pellet EVs. The pellet was resuspended in PBS and centrifuged furher at 120,000 × g for 70 minutes to enrich further EVs. Such EVs were subjected to SERS on the same day to measure the EV based spectra. Remaining EV samples were stored in the refrigerator at -80°C until used.

The EVs of 143B, Hela, and BMSC cells were selected for Transmission Electron Microscopy (TEM), Nanoparticle Tracking Analysis (NTA) characterization. TEM revealed teacup-shaped EVs, and NTA results showed that the diameters of the samples were concentrated around 100~150 nm. Western blotting was performed to compare the EV samples of 143B, Hela, and BMSC cells with the exosome surface proteins, CD9, and CD81 labelling, and the negative indicator, Calnexin protein. Protein electrophoresis comparison, TEM, NTA, and Western Blot results showed that we did obtain the exosomes of the above cells.

## 2.4 Collection of Sweat and Tear Samples

Three healthy volunteers were recruited for sweat sample collection in this study. Before collection, the forearm skin of each volunteer was cleaned with normal saline and dried. Sterile capillary were attached to the medial forearm skin, and sweat was collected by capillary force. (if this is a new tech should you tell this in more detail ? ) After collection, the sweat was immediately transferred into sterile centrifuge tubes via capillaries, centrifuged at low speed to remove skin debris and impurities, and stored at −20 °C for subsequent analysis.

This retrospective case-control study enrolled two patients with retinal vein occlusion, two with age-related macular degeneration, two with dry eye syndrome, two with diabetes mellitus (without diabetic retinopathy), two with diabetic retinopathy, two with pterygium, and two with hyperlipidemia, all from the Second Affiliated Hospital of Harbin Medical University. The study protocol was approved by the Ethics Committee of the School of Pharmacy, Harbin Medical University, and strictly adhered to the relevant provisions of the Declaration of Helsinki regarding human biomedical research. Tear samples were collected by placing a spotting capillary tube in the lateral canthal conjunctival sac of the patients. After collection, the tears were transferred from the capillary tube to a centrifuge tube using a capillary pipette and stored at -20°C for later use.

## 2.5 SERS substrate preparation and detection

Enhanced substrates were synthesized by the method we used previously, where 5 mL of silver nitrate solution (6.6 mg/ mL) was added to 495 mL of sodium borohydride solution (0.133 mg/ mL), stirred for 18 min, centrifuged at 2,870 ×g, 25°C for 20 min, and the supernatant was removed, and 10 μL of the silver nanosol mixed with 10 μL of the cellular EVs and 5 μL of sodium borohydride (0.05 M, pH=10) were mixed. The solution was stirred homogeneously and then subjected to SERS detection using a WITec Alpha 300 R (Ulm, Germany) instrument. The laser wavelength was 532 nm, the scan time was 30 s, the energy was 20 mW, and the cumulative number of times per test was one**.** Take 5 μL of silver nanosol, 5 μL of EVs, and 2.5 μL of sodium boroh dride solution (0.05 M, pH=10), mix them thoroughly with sufficient stirring, and then perform SERS detection. The SERS detection was conducted using a WITec Alpha 300R Raman spectrometer (Germany), with the detection parameters set as follows: laser wavelength of 532 nm, scanning time of 10 seconds, laser energy of 30 mW, and accumulation number of 1 time for each detection.

## 2.6 Data Processing and Machine Learning

We obtained 100 SERS spectra of each of these six cellular EVs, and the spectra were baseline corrected by LabSpec6 software, and the data range (600-1800 $cm^{-1}$) was selected for smoothing, and then the spectral data were normalized by Origin software to obtain the processed data.

PCA is a linear feature extraction technique, which projects the data into the low dimensional space by linear mapping, by doing so it can ensure that the original data has the highest variance in the low dimensional space. It does this by calculating the eigenvectors in the covariance matrix of its features. The eigen vectors corresponding to the largest eigen values (principal components) - are used to reconstruct new data and it is ensured that these data have the maximum variance in the direction of that eigen vector.

HCA is a cluster analysis method that aims to establish a hierarchical structure of clusters. It allows similar items to be grouped into clusters based on their characteristics and is an unsupervised learning algorithm. In hierarchical cluster analysis, data points are initially treated as individual clusters, which are then sequentially merged or divided based on their similarity until a single cluster containing all data points is formed. This process produces a hierarchical tree structure, called a dendrogram, which visually represents the relationships between clusters.

SVM is a class of generalized linear classifiers that binary classify data in a supervised learning fashion, and are linear classifiers defined to maximize the interval over the feature space, while the kernel trick also included in SVM makes him essentially nonlinear classifiers. Therefore, we choose SVM as the model for machine learning.

The processed spectral data was imported into RStudio and PCA downscaling was performed on the data using the prcomp function package in RStudio. SVM classifiers are built using the dimensionality-reduced data features PC1, PC2, PC3 and parameter search results to get the best classifier parameters, after building the best classifier model, 70% of the data (420 sets of spectra) will be used as a training set and 30% of the data (180 sets of spectra) will be used as a test set to test the accuracy of the model.

**Convolutional Neural Network (CNN)**-based classification was performed in the environment of Python 3.9+ and PyTorch 1.7, relying on the PyTorch, NumPy, Pandas, scikit-learn, Seaborn, and Matplotlib libraries. First, time-series feature data were prepared in CSV format and stratifiedly split into a training set (70%) and a test set (30%). The random seed was fixed at 123 to ensure reproducibility, and the batch size was set to 32. The model was constructed as a one-dimensional convolutional neural network (1D-CNN) using PyTorch, consisting of 2 Conv1d layers and 2 fully connected layers to output class probabilities. The model was trained for 500 epochs using the Adam optimizer (learning rate = 1e-3) and cross-entropy loss function.

**Recurrent Neural Network (RNN)**-based classification was conducted under the same environment as CNN (Python 3.9+, PyTorch 1.7, and the aforementioned dependent libraries). The dataset preparation process was consistent with that of CNN, including stratified splitting of the CSV-formatted time-series feature data into a 70% training set and a 30% test set, fixing the random_state at 123, and setting the batch size to 32. The model was constructed as a GRU-RNN classifier with 256 hidden units, and the hidden state was retained to achieve sequence memory. A threshold of $\theta$ = 1e-2 was set to prevent gradient explosion. The model was trained for 400 epochs using the Adam optimizer (learning rate = 1e-3) and cross-entropy loss function.

# 3. Results and discussion

## 3.1 Isolation and characterization of EVs

Ultracentrifugation is the gold standard for the isolation of EVs [42], including differential centrifugation and density gradient centrifugation, in which differential centrifugation does not require the addition of additional reagents and does not interfere with the process of label-free detection of EVs; therefore, we used differential centrifugation to purify and isolate EVs from cell lines after starvation culture, and the specific experimental procedure is shown in **Fig.2. a.**

The obtained EVs were characterized using conventional experiments such as TEM, NTA, and Western Blot. TEM results showed **(Fig.2. b)** that the EVs we obtained by differential centrifugation were between 100~150 nm in size and had no prominent protein aggregates. The particle size obtained by NTA **(Fig.2. d)** further confirmed the results (see **Fig. S1** for more information). Common EVs markers (CD9 and CD63) were selected as positive control proteins, and Calnexin protein was used as a negative control protein for Western blot **(Fig.2. c)**. These results indicated that we successfully obtained cellular sources EVs from cell culture fluid.

## 3.2 SERS detection of EVs

The enhanced substrate used for detecting cellular sources EVs was obtained using the method we have used before [43]. Ag@BO nanoparticles were obtained by reducing silver nitrate solution using sodium borohydride and centrifuging the solution. The specific detection process is shown in **Fig.1. a.** The dynamic light spectroscopy results of the mixed solution of each part are shown in **Fig.S2**, where there was no Raman signal for both Ag@BO nanoparticles and EVs. We mixed Ag@BO nanoparticles with cellular EVs, and still no signal was detected, as shown in **Fig.2. b** Ag@BO nanoparticles were not tightly bound together after mixing with EVs. When we add sodium borohydride solution, the reducing property of sodium borohydride keeps the surface of the nanoparticles clean and acts as an aggregator causing Ag@BO

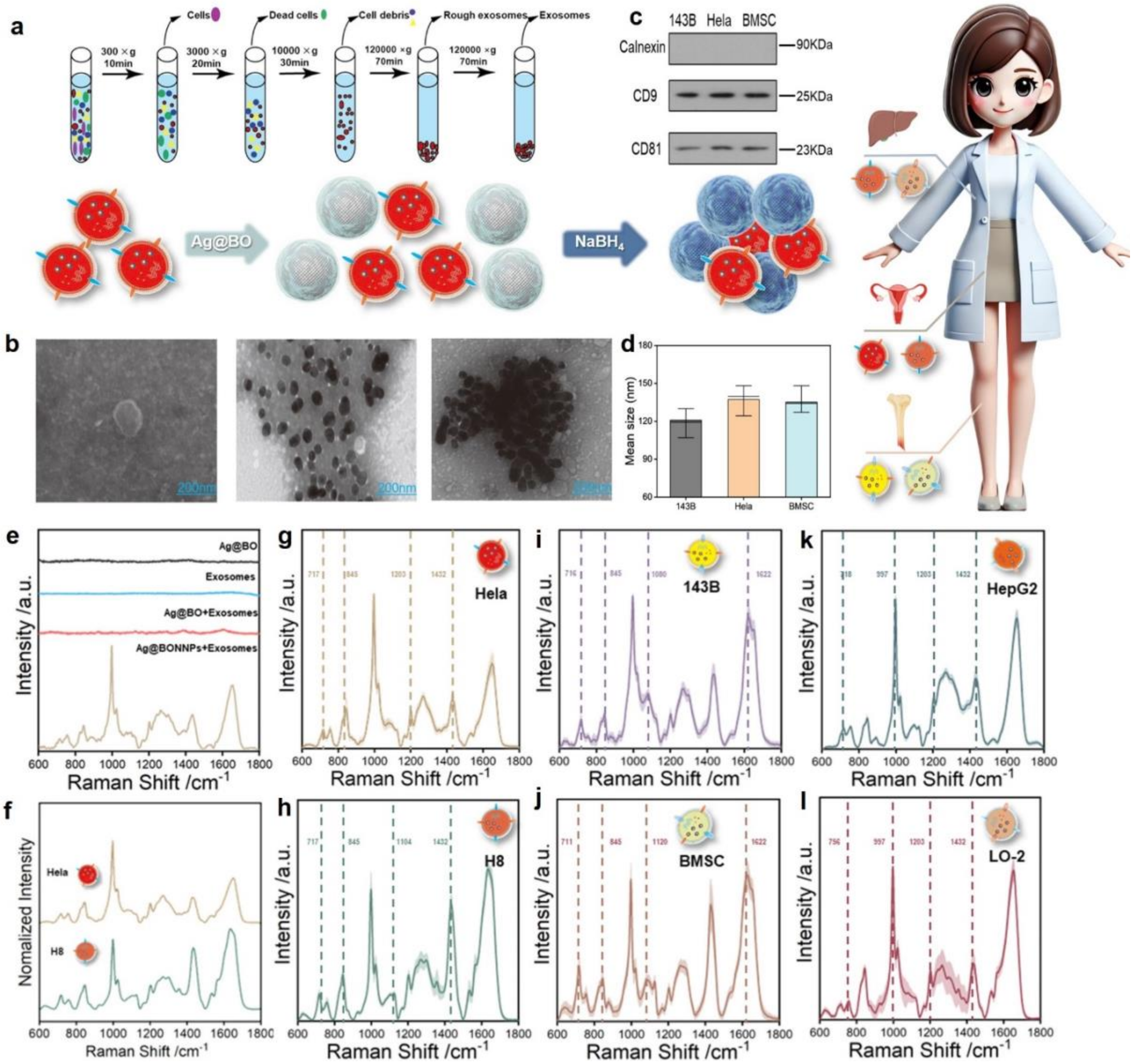

*Figure 2*: *Isolation, characterization, and SERS profiles of cell-derived EVs. a. Isolation and extraction process of cell-derived EVs. b. TEM of cell-derived EVs, EVs with Ag@BO NPs, and EVs with Ag@BONNPs, scale bar: 200 nm. c-d. Western Blot and NTA results of the three cell-derived EVs of 143B, Hela, BMSC. e. SERS profiles of Hela cell-derived EVs under different systems. f. SERS profiles of Hela cell and H8 cell-derived EVs. j-l. SERS profiles of six different cell-derived EVs.*

nanoparticles to bind to the amino group on the surface of the EVs to form silver-ammonia covalent bonds, Ag@BO nanoparticles were adsorbed on the EV surface, like a “cleanser”, we were able to get more material information about the EVs, and therefore we were able to get the signals of the cellular sources EVs **(Fig.2. e)**. The 15 sets of SERS profiles of the EVs of Hela cells detected at different times had an RSD of 5.2% at the peak position of 845 $cm^{-1}$ with good reproducibility **(Fig.S3. a)**. Subsequently, we examined EVs from H8 cells and found that their SERS signals were very similar to but still distinct from those of Hela cell-secreted EVs **(Fig.2. f)**, demonstrating good specificity of the selected method. We then examined and mapped the cellular EVs of HepG2, 143B, LO-2 and BMSC using the method used earlier here in. 100 spectra were obtained from each cell-derived EVs. These signals were preprocessed, including baseline correction, denoising, and min-max normalization, and peak attribution was performed (see **Supplementary Information Table 1** for details); the average spectra obtained from the processing of 100 spectral data for each cell are shown in **Fig.2. g-l**. Since these cellular sources EVs have the same composition and differ only slightly in content, resulting in slight differences in their SERS spectra, it is difficult for the human eye to observe such minor differences. Given this we went on to use thermogram, PCA and HCA methods to demonstrate putative differences in the main features of spectral data of these cellular source's EVs.

To gain insight into the dynamic association between EVs proteins and silver nanoparticles (AgNPs), molecular dynamics simulations were performed, and snapshots representing the exact state of the system at specific time points were extracted for detailed analysis. **Fig. 3.a–k** display a series of these snapshots for the exosome–AgNP mixture, visually capturing the evolution of the system over the course of the simulation. These images can be interpreted to illustrate the progressive adsorption of the protein onto the nanoparticle surface.

The structural stability of the protein–AgNP complex was quantitatively assessed by the root-mean-square deviation (RMSD) of the protein backbone atoms over a 100 ns trajectory. As shown in **Fig. 3.l**, the RMSD value gradually plateaued and reached a final value of 1.29 nm at the end of the simulation, indicating that the complex attained a stable conformation during the last portion of the trajectory.

The energetics of the binding process were further examined by decomposing the non-bonded interaction energy between the protein and the silver nanoparticle. **Fig. 3.m** presents the time evolution of the short-range Lennard–Jones (LJ–SR) and short-range Coulombic (Coulomb–SR) contributions. The LJ–SR energy exhibited a strongly favorable profile, with a final value of –1481.22 kJ/mol and an average value of –987.94 kJ/mol over the trajectory. In contrast, the average Coulomb–SR energy remained 0 kJ/mol throughout the simulation. Under the force field employed here, silver atoms are typically modeled as neutral particles; therefore, the direct electrostatic term is essentially null. Nevertheless, this does not preclude the formation of specific polar contacts, as the large favorable van der Waals component, together with the geometric complementarity observed, drives the tight binding.

The molecular basis of this affinity is addressed in **Fig. 3.n** that provides a close-up view of the protein–Ag interface at the end of the simulation. Multiple oxygen atoms from the protein backbone and side chains were found in close proximity to silver atoms, with interaction distances ranging from 2.74 to 3.24 Å, a range indicative of strong, specific binding. The residues that were involved in these contacts were Asp96, Leu94, Gly86, Ala53, Ala82, Asn74, Gln54, and Glu85. Specifically, the silver atoms interacted predominantly with backbone carbonyl oxygens as well as with carboxylate oxygens of the side chains (such as OD1 of Asp/Asn and OE1 of Glu/Gln). These polar atoms acted in the simulations as multiple anchor points that tethered the protein to the AgNP surface through concerted, multi-site contacts. Rather than relying on long-range electrostatics, the stabilization seemed to arise according to the computing data from the dense network of oxygen–Ag interactions, which are essentially coordination-like contacts that contributed significantly to the van der Waals and induced-polarization terms in the energy description.

Collectively, these simulation modelling results raise the point that the EV protein recognizes and binds to the silver nanoparticle via a multi-dentate mechanism. The surface-exposed oxygen-containing functional groups—mainly carbonyl and carboxylate moieties—serve as discrete anchoring sites, collectively establishing a robust and stable protein–silver interfacial architecture. This interaction mode explains the substantial van der Waals stabilization observed and underlines why the protein remains firmly adsorbed even in the absence of net Coulombic attraction, a finding consistent with the behavior of proteins on noble metal surfaces where dispersive and charge-transfer interactions often dominate.

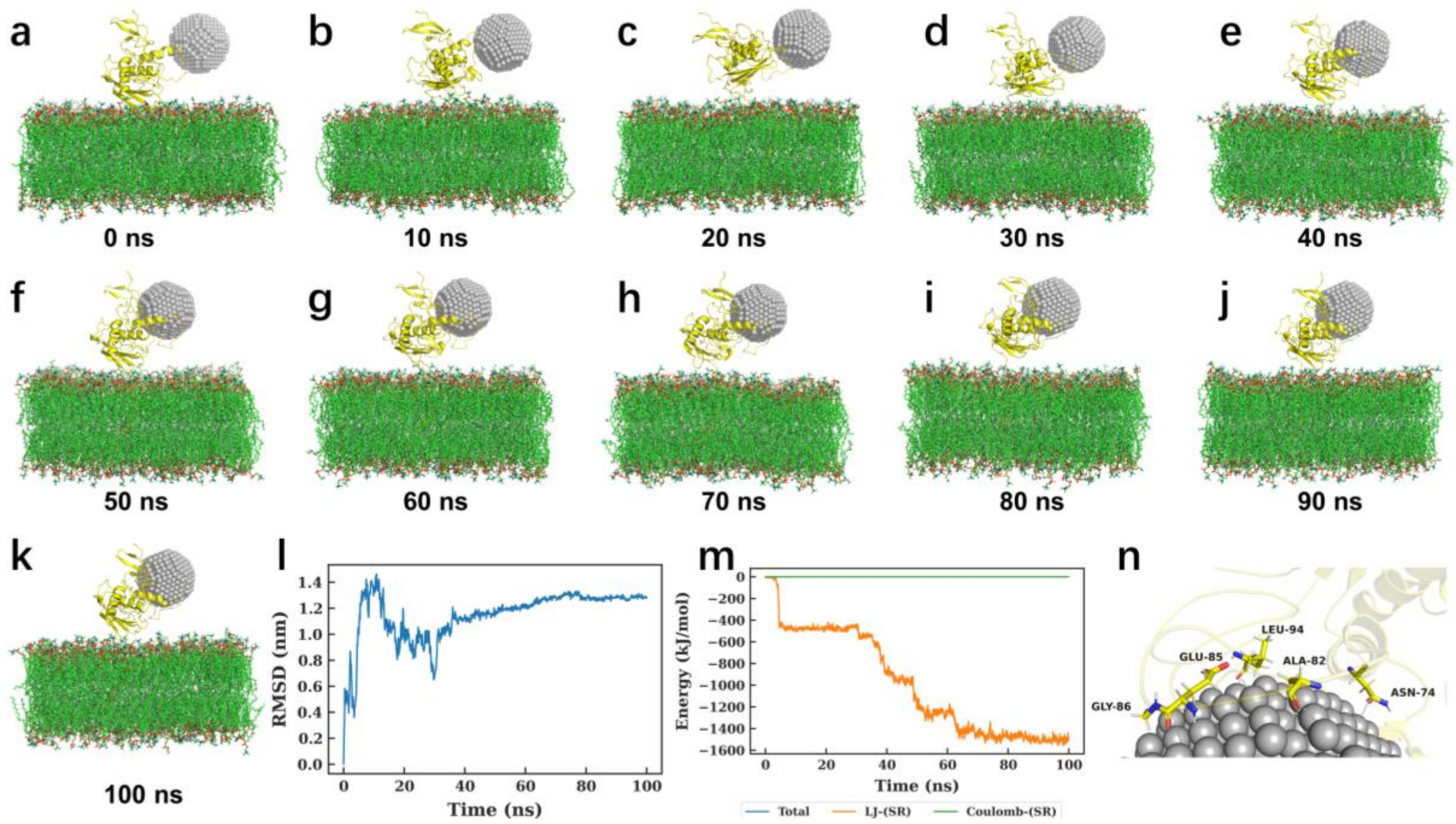

*.Figure 3. Molecular dynamics simulation of the EV-silver nanoparticle (AgNP) system. (a–k) Representative snapshots of the EV protein-AgNP system at different simulation times (0–100 ns). (l) Time-dependent root-mean-square deviation (RMSD) curve of the EV surface protein-AgNP complex. (m) Time evolution of van der Waals (LJ-SR) and electrostatic (Coulomb-SR) interaction energies between the EV surface protein and AgNPs. (n) Detailed view of the binding interactions between key amino acid residues of the EV protein and the AgNP surface.*

## 3.3 Thermograms, PCA and HCA of SERS profiles of EVs from cancer and normal cell sources

We first performed a thermogram comparison, PCA and HCA on the SERS profiles of three different cancer cells and their corresponding normal cells, in which the EV signals originated from Hela cells and H8 cells could be well distinguished, with notable differences **(Fig.4. c-e)**. Differences in EV signals between HepG2 cells and LO-2 cells were minute, heatmaps did not differ, a nd moreover the results of hierarchical clustering analysis were not conclusive either. However the two cases could be differentiated by PCA **(Fig.4. f-h)**. Also the thermogram comparisons of EV signals beteen the 143B cells and the BMSC cells depicted differences. Indeed the 95% of the PCA confidence intervals overlapped partially, but the HCA algoritm distinguished the two smaples **(Fig.4. i-k)**. Based on these initial data we went on to to compare the signals obtained from either of the model cancer cells and those of normal cells serving as controls. The highest SERS raman peaks of normal cellular sources EVs were obtained from the 1654 $cm^{-1}$ peak of amide I, i.e., the C=C stretching vibration. In contrast, cancer cell derived EVs provided SERS Raman peaks at the 997 $cm^{-1}$ at the region of symmetric respiratory vibration of phenylalanine. This feature will be useful when studying the cancer cells in more in detail.

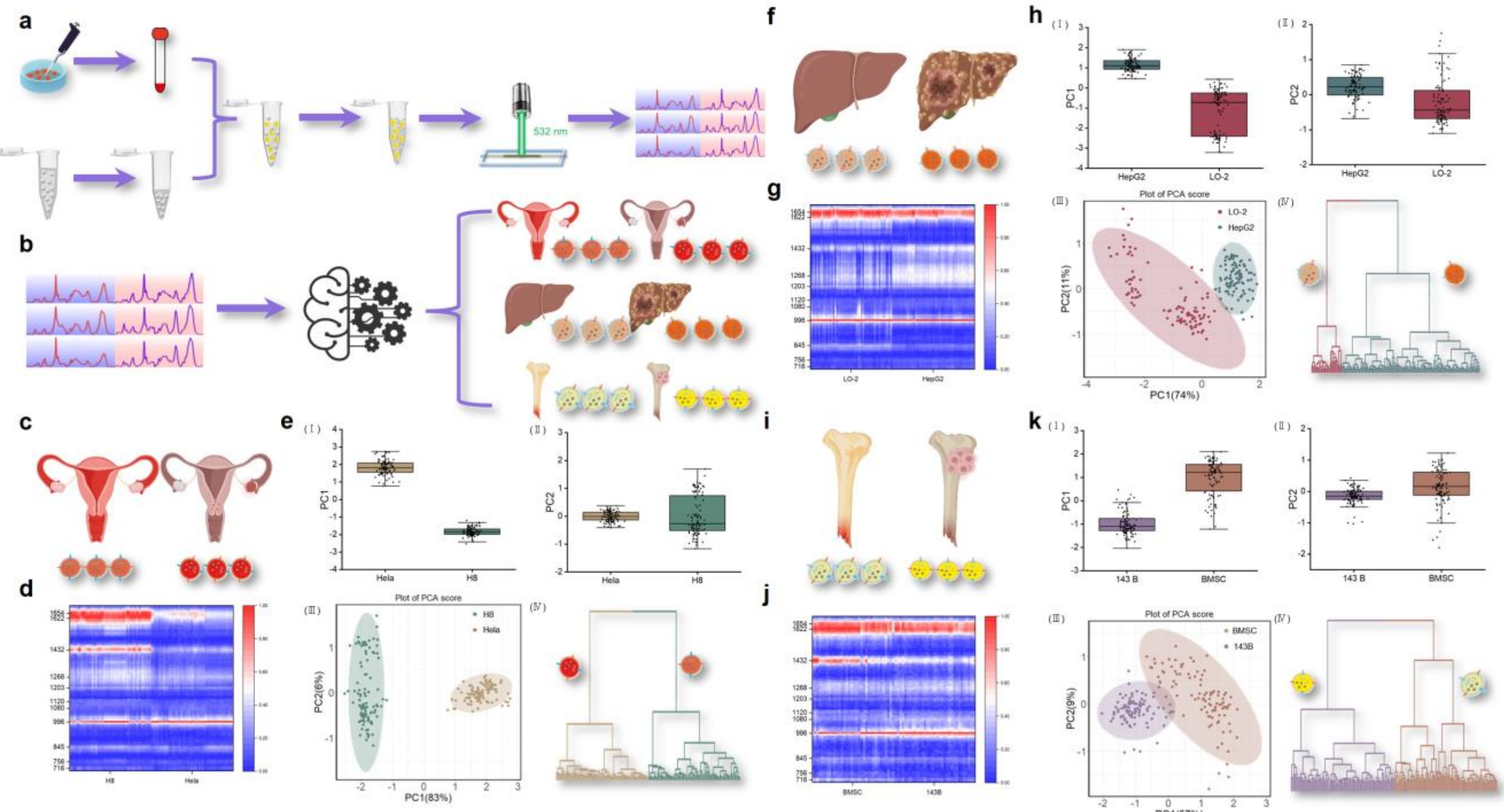

*Figure 3: Detection flow chart, Thermograms, Principal component analysis, and Hierarchical clustering analysis of exosomes from three cancer cells and their corresponding normal cell sources. a-b. Diagram of the process of detecting and distinguishing cell-derived EVs between cancer and normal cell-derived EVs. Schematic diagrams of cervical cancer(c), hepatocellular carcinoma(f) and osteosarcoma(i). Thermograms comparison of EVs derived from Hela cells and H8 cells(d), HepG2 cells and LO-2 cells(g), 143B cells and BMSC cells(j). PCA and HCA of EVs derived from Hela cells and H8 cells (e), HepG2 cells and LO-2 cells (h), 143B cells and BMSC cells (k). Cartoons were created with BioRender.com.*

To investigate the differences in EVs from different cellular sources, we used thermograms to demonstrate the spectral signatures of cancer cells from three different tissue sources (143B, Hela and HepG2) **(Fig.5. b)**, and then computed the eigen vectors in the spectral eigen-covariance matrices using unsupervised dimensionality reduction using PCA and projected the data into a low-dimensional space by linear mapping **(Fig.5. c.III)**. The cumulative contribution of PC1 and PC2 reaches 86%, indicating that the first two principal components explain the variability of the spectral data well. We were able to find the principal component differences in the SERS spectra of EVs from cancer cells of different tissue origins by PCA **(Fig.5. c.I-II)**, and in this way, we were able to distinguish them. To further highlight the differences between EVs from different cellular sources, we selected HCA to analyse the SERS data. The results of HCA based on Euclidean distance were the same as those of PCA, and the SERS profiles of EVs from the three cancer cell sources were classified into three categories with significant differences **(Fig.5. c.IV)**.

To exclude the specificity of the appearance of differences in EVs of cancer cell origin, we also demonstrated the EV signals of the three normal cell origins using thermograms, PCA and HCA **(Fig.5. d-f)**, and the results of the PCA showed that the cumulative contribution of PC1 and PC2 was 73%. Even though the cumulative contribution was slightly lower than that of the results of the EVs of cancer cell origin, they could still be distinguished from each other.

The results of PCA were further verified by HCA **(Fig.5. f.Ⅳ)**. The results showed that there were also significant differences in EV signals from different normal cellular sources, the signal differences of EVs can be shown more intuitively by thermograms, PCA, and HCA. However, it cannot completely and accurately distinguish the EVs signals of cancer cell sources and normal cell sources, we tried to differentiate the SERS profiles of different cellular EVs using an SVM model, and we need used PCA to downsize the data to avoid the appearance of overfitting because the SERS data contains a large number of data features. So we construct a PCA-SVM combined machine learning model to distinguish EV signals from different cell sources and make predictions.

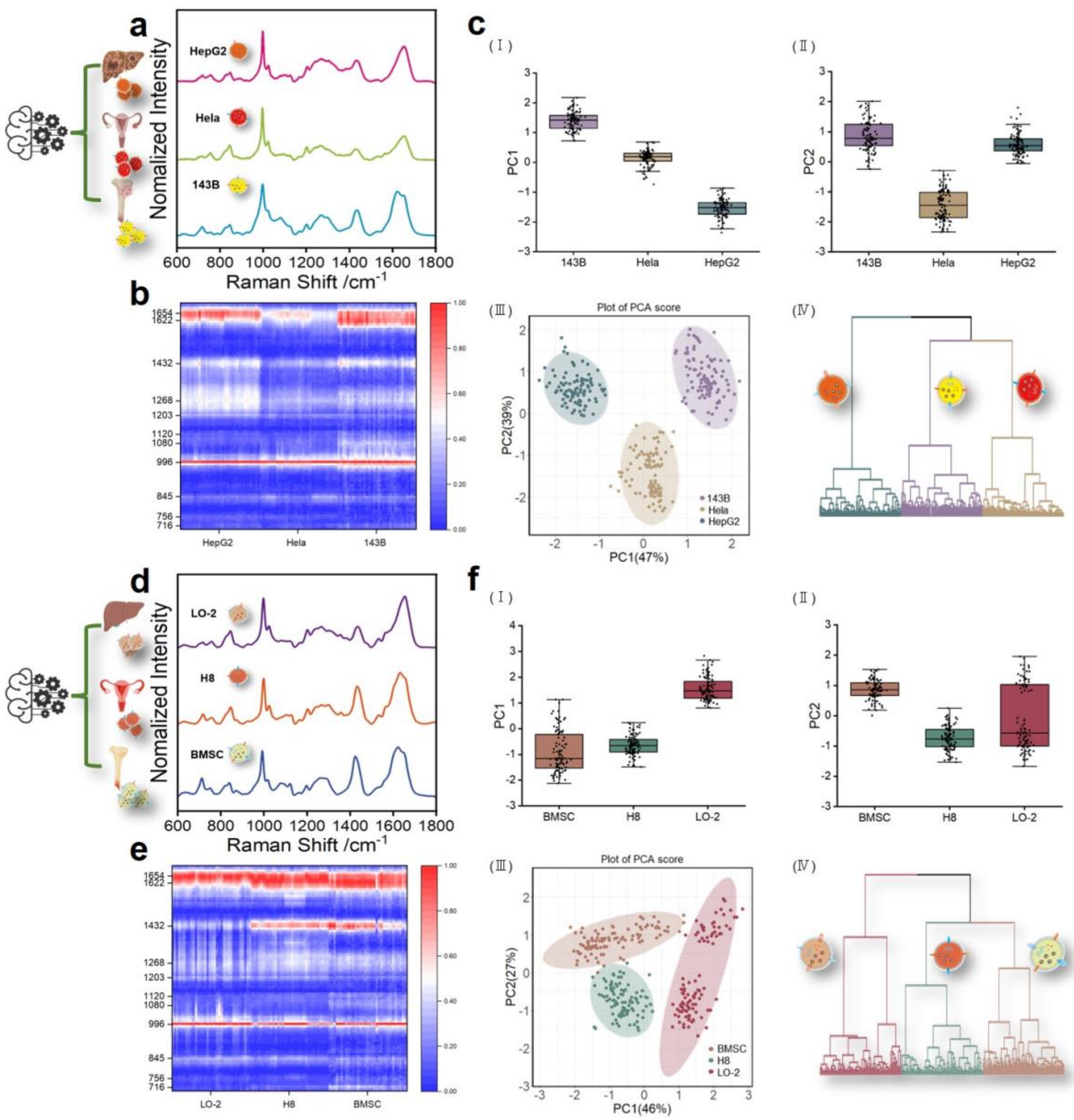

*Figure 4*：*SERS profiles, Thermograms, PCA, and HCA of three normal cell-derived EVs and three cancer cell-derived EVs. a-b. SERS profiles of EVs and thermograms from three cancer cell sources. d-e. SERS profiles of EVs and thermograms from three normal cell sources. PCA and HCA results of EVs from three cancer cell sources(c) and three normal cell sources(f),(I-III) the results of PCA,(IV) the results of HCA.*

## 3.4 Classification and prediction of different cellular sources and ophthalmic diseases sources EVs using machine learning

Initially, we found that the conventional 2D PCA could not well distinguish the signals of the six cellular EVs with most of the overlapping parts **(Fig.S3. b).** Therefore, we adopted a three-dimensional PCA **(Fig.6. b)** and formed a feature dataset by calculating the PC1, PC2 and PC3 eigen values of each cellular EVs. We divided this dataset into a 70% training set and a 30% test set and then used the training set to train the SVM model and the test set to validate the performance of the model. The classification interface of the SVM model is shown in **Fig.S4**. Finally, we used the test set to evaluate the model's performance and generated the confusion matrix **(Fig.6. c)**.

The detailed results of the confusion matrix can be viewed in **Supplementary Information Table 2.** According to the confusion matrix results, our model achieves an accuracy of 94.4%. This indicates that our method can effectively detect the SERS profiles of EVs of different cellular origins and realize their differentiation and prediction by combining the PCA-SVM machine-learning model.

To evaluate the practical applicability of the proposed SERS-based biosensing platform for non-invasive biofluid analysis, sweat samples were collected from three healthy volunteers (**Fig.6. d**). The corresponding Raman spectra, measured under identical instrumental conditions, are presented in **Fig.6. e**. All three sweat samples exhibited similar spectral features in the 600–1800 $cm^{-1}$ region, including characteristic bands attributed to major sweat constituents such as EVs, lactate, urea, and amino acids. Nevertheless, consistent inter-individual variations in relative peak intensities were observed, reflecting the biochemical heterogeneity among volunteers.

To further resolve these inter-individual spectral differences, PCA was performed on the Raman dataset. As shown in **Fig.6. f**, PC1 and PC2 accounted for 81.4% and 7.7% of the total variance, respectively. The resulting PCA score plot revealed three well-separated clusters corresponding to Sweat 1, Sweat 2, and Sweat 3, with no overlap between groups. This clear segregation demonstrates that sweat Raman spectra carry sufficient individual-specific information to enable reliable differentiation among subjects.

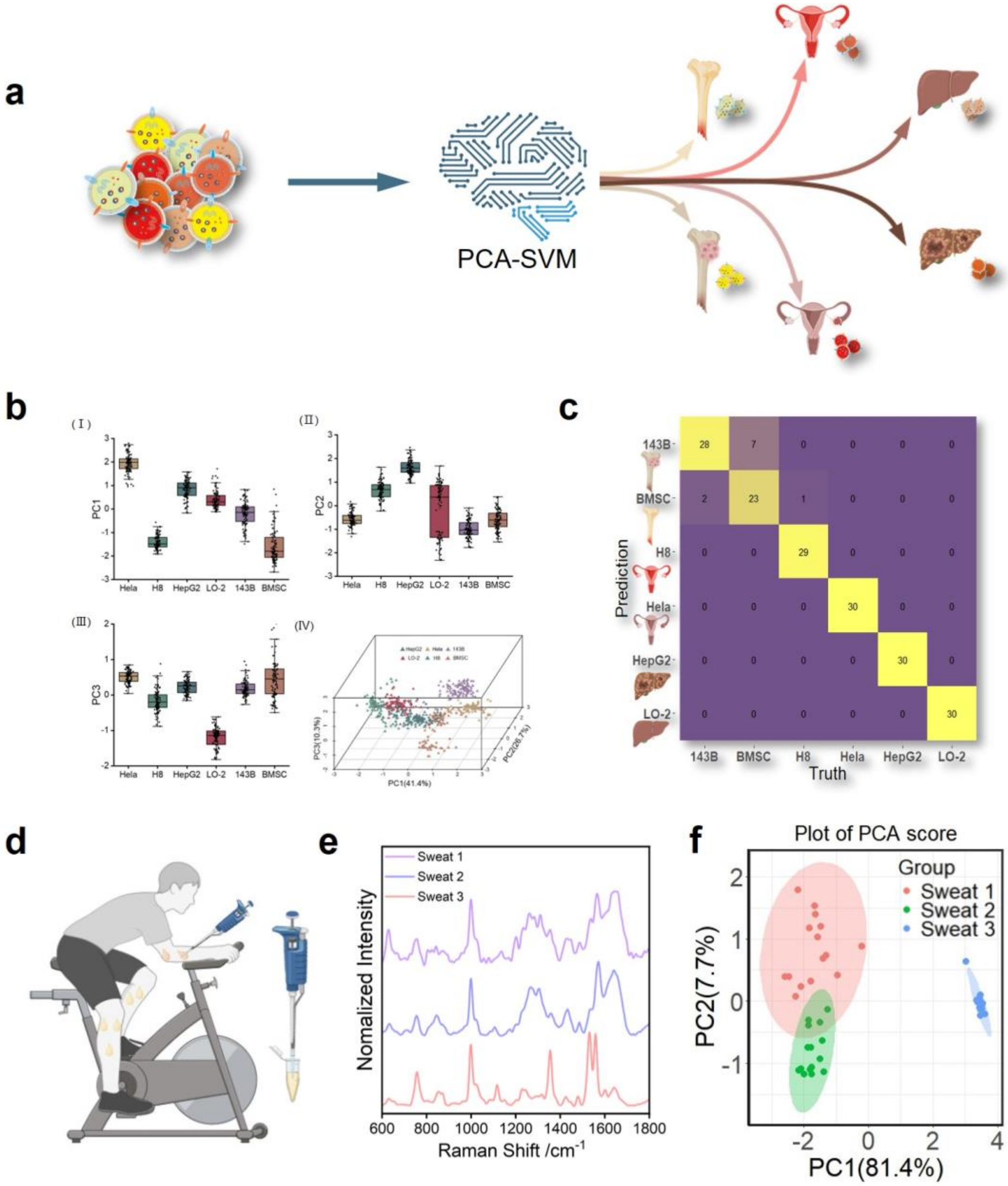

*Figure 6 : 3D PCA of six different cell-derived EVs and confusion matrix obtained from machine learning results. a. Conceptual diagram of the PCA-SVM machine learning algorithm to distinguish six different cell-derived EVs. b. (I-III). Comparison of PCA scores for EVs from six different cellular origins. (IV)3D PCA of SERS profiles of EVs from six cell-derived EVs. (PC1=41.4%, PC2=26.7%, PC3=10.3%). c. Prediction results of the three principal components (PC1, PC2, PC3) scores of six cell-derived EV SERS profiles on the test set. d. Schematic illustration of simulated sweat collection from a volunteer. e. Raman spectra of sweat samples obtained from three volunteers using the method described in this study. f. PCA results of the Raman spectra of sweat samples from the three volunteers.*

To explore the molecular fingerprints of tears associated with various ocular pathologies, we collected tear samples from patients with age-related macular degeneration (AMD), diabetic retinopathy (DR), diabetic macular edema (DME), retinal vein occlusion (RVO), dry eye syndrome (DE), hyperlipidemia-related ocular complications (HL), pterygium (PT), and healthy controls (NC). The representative SERS spectra acquired from these samples are presented in **Fig.7.b**, showing distinct intensity and peak distribution patterns across different groups. Notable spectral

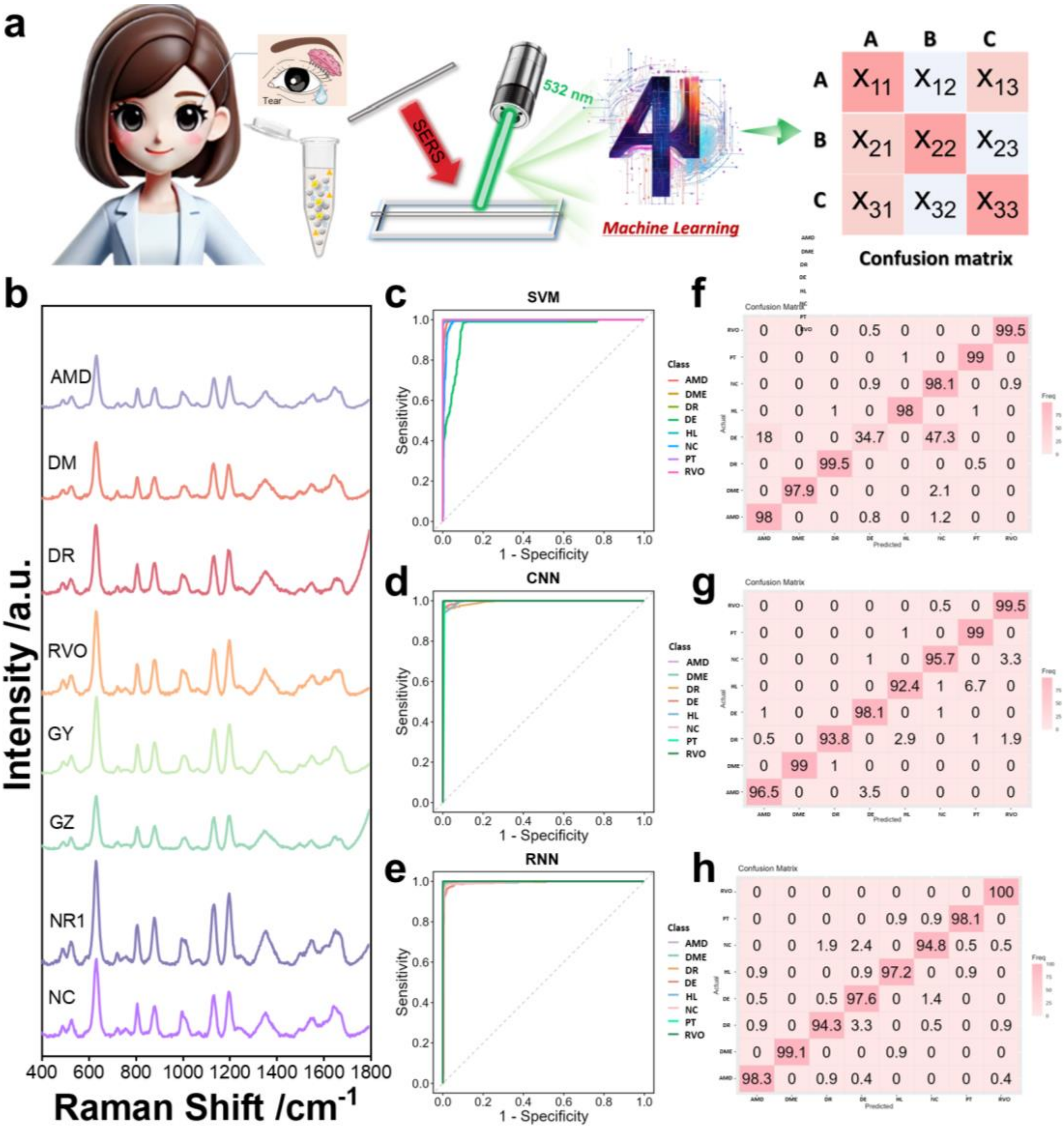

*Figure 7: a. Workflow schematic depicting tear collection from patients with ocular diseases, SERS-based spectroscopic analysis, and subsequent machine learning-aided classification. b. Typical SERS spectra acquired from tear samples of patients with seven distinct ocular diseases and healthy volunteers. c-e. Receiver operating characteristic (ROC) curves demonstrating the performance of three machine learning algorithms (SVM, CNN, and RNN) in distinguishing different ocular conditions. f-h. Confusion matrices illustrating the classification performance of the SVM, CNN, and RNN models, respectively.*

differences were observed in the regions corresponding to characteristic biomolecular vibrations.

To quantitatively classify the tear SERS spectra, three machine learning models-support vector machine (SVM), convolutional neural network (CNN), and recurrent neural network (RNN) were implemented.

Three machine learning models (SVM, CNN, RNN) were evaluated for multi-class classification. As shown in **Fig. S5**, the CNN model achieved the highest accuracy (0.972), followed by RNN (0.969) and SVM (0.884). ROC curves (**Fig.7.c–e**) and confusion matrices (**Fig.7.f–h**) confirmed that deep learning models outperformed traditional SVM, likely due to their superior ability to capture subtle spectral features. This SERS-machine learning platform thus demonstrates great potential as a rapid, non-invasive tool for objective, multi-class ocular disease screening, supporting timely intervention and point-of-care testing applications.

# 4. Conclusions

In conclusion, we developed a label-free method for rapid detection of EVs from different cellular sources and biological fluid-derived samples. We combined it with the PCA-SVM model to classify and predict EVs from different cellular sources, and the accuracy of identifying EVs from six cellular sources reached 94.4%. Building on this foundation, we further extended the platform to analyze tear samples from eight distinct ocular conditions (including seven types of eye diseases and healthy controls) and sweat samples from three healthy volunteers.

Our comparative evaluation of three machine learning classifiers revealed that deep learning models (CNN, 97.2%; RNN, 96.9%) significantly outperformed the traditional SVM (88.4%) in differentiating tear SERS spectra, highlighting the superior ability of deep learning to capture subtle spectral features associated with ocular pathologies. The exosome detection method we developed by combining SERS and artificial intelligence has the following advantages: (i) The method allows rapid label-free detection of detailed fingerprint profiles of EVs with a high signal-to-noise ratio compared to other methods; (ii) The fingerprints obtained reflect the difference in EV composition between cancer cells and normal cells;(iii) We were able to use only 10 μL of sample for the detection of EV samples and accurately differentiate EVs from different cellular sources with 94.4% accuracy after training with a small amount of data. Overall, our study presents a versatile, multi-functional platform for analyzing biological fluid-derived EVs and tear/sweat samples. It not only provides a simple and rapid tool for investigating EV composition, intercellular communication, and disease mechanisms but also offers a new avenue for the development of targeted EV-based therapies. Furthermore, the successful extension of the platform to ocular disease classification underscores its potential as a non-invasive, point-of-care diagnostic tool for multi-class disease screening, paving the way for future applications in precision medicine.

## CRediT authorship contribution statement

**Yang Li, Xiaoming Lyu:** Experimental part, data processing, artificial intelligence, drawing charts, first draft writing; **Xia Ling**: data processing and manuscript editing. **Haoyu Ji:** cell culture, providing exosome samples and exosome characterisation experiments; **Lei Qin:** cell culture, Extraction of cellular exosomes; **Kuo Zhan:** manuscript editing; **Seppo Vainio:** conceptualization and manuscript editing**; Jian-An Huang:** Conceptualization, Methodology, Writing – review & editing, Supervision, Project administration.

## Acknowledgments

This work was supported by the National Natural Science Foundation for Youth under Grant (No. 82202648); the DigiHealth project under Grant (project number 326291), a strategic profiling project at the University of Oulu that is supported by the GeneCellNano flagship of sAcademy of Finland and the University of Oulu.